\documentclass[%
 reprint,
 amsmath,amssymb,
 aps,
]{revtex4-2}
\PassOptionsToPackage{numbers,sort&compress}{natbib}

\usepackage{subfigure,dcolumn}
\usepackage[T2A,T1]{fontenc}
\usepackage[russian,english]{babel}

\usepackage{listings}
\usepackage{amsthm,amsmath,amssymb}
\usepackage{mathrsfs}
\usepackage{dutchcal}
\usepackage{siunitx}

\usepackage{tikz,xcolor,hyperref}

\definecolor{lime}{HTML}{A6CE39}
\DeclareRobustCommand{\orcidicon}{
	\begin{tikzpicture}
	\draw[lime, fill=lime] (0,0) 
	circle [radius=0.16] 
	node[white] {{\fontfamily{qag}\selectfont \tiny ID}};
	\draw[white, fill=white] (-0.0625,0.095) 
	circle [radius=0.007];
	\end{tikzpicture}
	\hspace{-2mm}
}
\foreach \x in {A, ..., Z}{%
	\expandafter\xdef\csname orcid\x\endcsname{\noexpand\href{https://orcid.org/\csname orcidauthor\x\endcsname}{\noexpand\orcidicon}}
}

\usepackage[numbers,sort&compress]{natbib}
\begin{document}

\title{Reduced Basis Method for Few-body Bound State Emulation}

\thanks{Discussions with Pablo Giuliani and Witold Nazarewicz are gratefully acknowledged. The work was partially supported by the National Key Research and Development Program (MOST 2022YFA1602303 and MOST 2023YFA1606404), the National Natural Science Foundation of China under Contract No.\,12347106 and No.\,12147101.}

\author{R. Y. Cheng}
\affiliation{Key Laboratory of Nuclear Physics and Ion-beam Application (MOE), Institute of Modern Physics, Fudan University, Shanghai 200433, China}
\affiliation{Shanghai Research Center for Theoretical Nuclear Physics, NSFC and Fudan University, Shanghai 200438, China}

\author{K. Godbey}
\affiliation{FRIB/NSCL Laboratory, Michigan State University, East Lansing, MI, United States}

\author{Y. B. Niu}
\affiliation{Key Laboratory of Nuclear Physics and Ion-beam Application (MOE), Institute of Modern Physics, Fudan University, Shanghai 200433, China}
\affiliation{Shanghai Research Center for Theoretical Nuclear Physics, NSFC and Fudan University, Shanghai 200438, China}

\author{Y. G. Ma\orcidD{}}
\affiliation{Key Laboratory of Nuclear Physics and Ion-beam Application (MOE), Institute of Modern Physics, Fudan University, Shanghai 200433, China}
\affiliation{Shanghai Research Center for Theoretical Nuclear Physics,
NSFC and Fudan University, Shanghai 200438, China}

\author{W. B. He}
\affiliation{Key Laboratory of Nuclear Physics and Ion-beam Application (MOE), Institute of Modern Physics, Fudan University, Shanghai 200433, China}
\affiliation{Shanghai Research Center for Theoretical Nuclear Physics,
NSFC and Fudan University, Shanghai 200438, China}

\author{S. M. Wang\orcidE{}}
\email[Corresponding author, ]{wangsimin@fudan.edu.cn}
\affiliation{Key Laboratory of Nuclear Physics and Ion-beam Application (MOE), Institute of Modern Physics, Fudan University, Shanghai 200433, China}
\affiliation{Shanghai Research Center for Theoretical Nuclear Physics,
NSFC and Fudan University, Shanghai 200438, China}

% \affil[1]{Key Laboratory of Nuclear Physics and Ion-beam Application (MOE), Institute of Modern Physics, Fudan University, Shanghai 200433, China}
% \affil[2]{Shanghai Research Center for Theoretical Nuclear Physics,
% NSFC and Fudan University, Shanghai 200438, China}
% \affil[3]{FRIB/NSCL Laboratory, Michigan State University, East Lansing, MI, United States}

\begin{abstract}
% New computational methods are urgently needed in nuclear physics to address the inefficiencies and limitations of current theoretical models, which often struggle with expanding model spaces and high time consumption.
% This study presents a novel reduced basis method (RBM) designed to substantially lower computational costs by constructing a significantly smaller Hamiltonian subspace informed by previous solutions. It offers a flexible alternative to the eigenvector continuation (EC) method, which has been widely used in nuclear physics recently. Our method shows comparable efficiency and accuracy to EC on an artificial three-body bound system while providing a richer representation of physical information in its projection and training subspace. This methodological advancement has the potential to greatly improve computational techniques and enhance our understanding of the fundamental components of nuclear systems.

Recent advances in both theoretical and computational methods have enabled large-scale, precision calculations of the properties of atomic nuclei.
With the growing complexity of modern nuclear theory, however, also comes the need for novel methods to perform systematic studies and quantify the uncertainties of models when confronted with experimental data.
% One growing sub-field
This study presents an application of such an approach, the reduced basis method, to substantially lower computational costs by constructing a significantly smaller Hamiltonian subspace informed by previous solutions.
Our method shows comparable efficiency and accuracy to other dimensionality reduction techniques on an artificial three-body bound system while providing a richer representation of physical information in its projection and training subspace.
This methodological advancement can be applied in other contexts and has the potential to greatly improve our ability to systematically explore theoretical models and thus enhance our understanding of the fundamental properties of nuclear systems.
\end{abstract}

\keywords{reduced basis method, Gamow coupled-channel, few-body, emulator}

\maketitle 

%%%%%%%%%%%%%%%%%%%%%%%%%%%%%%%%%%%%%%%%%%%%%%%%%%%%%%%%%%%%%%%%%%%%%%%%%%%%%%%%%%%%%%%%%%%%%%
%%%%%%%%%%%%%%%%%%%%%%%%%%%%%%%%%%%%%% INTRODUCTION %%%%%%%%%%%%%%%%%%%%%%%%%%%%%%%%%%%%%%%%%%
%%%%%%%%%%%%%%%%%%%%%%%%%%%%%%%%%%%%%%%%%%%%%%%%%%%%%%%%%%%%%%%%%%%%%%%%%%%%%%%%%%%%%%%%%%%%%%

\section{Introduction}

Exploration of new computational methods is at the forefront of advancements in modern nuclear physics~\cite{Carlson2015a,Navratil2016a}. As the number of nucleons increases, the complexity and dimensionality of the system within the nucleus also grow exponentially, significantly challenging the computational capabilities and questioning the adequacy of traditional methodologies. Among these, {\it ab initio} approaches are considered to be the most accurate and reliable frameworks in nuclear physics. However, despite notable advances in recent years~\cite{Hergert2016a}, only a few models manage to converge without appropriate truncations. This restricts the ability of the models to investigate the more enigmatic phenomena associated with exotic nuclei, particularly those near the drip line. These nuclei exhibit strong coupling with their environment, and their behavior is heavily influenced by continuum effects. Incorporating these effects not only complicates the computational problem but also substantially increases the associated computational costs~\cite{Michel2021a,Li2021a,Michel2002a}. This complexity poses significant challenges to theoretical predictions and necessitates extensive computational resources for systematic calculations.

Recent developments in machine learning \cite{Boehnlein2022a,He2023b,Shang2022a} and statistic \cite{Alhassan2022a} methods present an alternative way for addressing the computational challenges inherent in complex physics problems by facilitating the extraction of principal components. These advancements significantly enhance computational efficiency and, in some cases, provide reliable predictions. Machine learning methods applicable to nuclear physics can be broadly categorized into data-driven and model-driven approaches \cite{Brunton2019a,He2019a}. These methodologies have been progressively introduced to the nuclear physics community to mitigate computational burdens \cite{Boehnlein2022a,He2023b,Ma2023a,He2023a,Wang2020a}.
Data-driven machine learning algorithms, such as artificial neural networks, Gaussian processes, and Bayesian neural networks, have demonstrated significant progress in the field of nuclear physics~\cite{He2021a,Utama2016a,Neufcourt2020a,Neufcourt2020b,Huang2022a}. These algorithms are particularly effective in analyzing extensive experimental datasets and supporting theoretical predictions. For instance, Bayesian methodologies have been applied to model mixing~\cite{Kejzlar2023a, Semposki2022a,Pang2023a} and uncertainty quantification~\cite{Odell2024a, Smith2024a, Giuliani2023a, Giuliani2024a}. This application facilitates the rigorous calibration of microscopic nuclear models by determining the posterior distributions of critical physical parameters such as potentials and masses.
On the other hand, emerging methods inspired by physical principles are aiding theorists in refining their models, reducing computational demands, and deepening insights into specific nuclear phenomena. These model-driven approaches integrate physical laws directly into the learning algorithms, providing a more robust framework for tackling the multifaceted challenges in nuclear physics.

Among these methodologies, Eigenvector Continuation (EC) stands out as a recently proposed novel method~\cite{Bai2021a,ZhangX2024a,Yapa2023a,Yapa2024a}. It is founded on the principle that the wave function changes smoothly with adjustable parameters in the Hamiltonian, a concept that has been demonstrated to be effective in solving eigenvalue problems~\cite{Frame2018a,Frame2019a,Sarkar2022a,Sarkar2021a}. The EC method exemplifies a specific application of the broader Reduced Basis Method (RBM), which encompasses a range of numerical techniques designed for rapid and accurate simulations of parametric systems through strategic training and projection \cite{Melendez2022a, Drischler2023a, Quarteroni2016a, Bonilla2022a, Duguet2024a}. 
Distinct from other data-driven machine learning approaches~\cite{Brunton2019a,Kutz2016a}, RBM adopts a model-driven strategy that aligns with the intrinsic nature of the underlying physical model. This approach operates within a high-fidelity framework, where the Hamiltonian is meticulously constructed via a microscopic physical model, facilitating thorough offline preparations for subsequent simulations~\cite{Drischler2023a}. During the so-called online stage, the dimensionality of the problems under consideration is effectively reduced. This reduction is achieved by projecting the high-fidelity model onto a significantly smaller subspace, offering a strategic alternative to merely training datasets without yielding interpretability. This methodology enhances the computational efficiency while maintaining a rigorous connection to the physical underpinnings of the problem.

In RBMs, the selection of appropriate training and projecting subspaces is pivotal for the specific system under investigation. The Galerkin method, a widely used projection technique in RBM, generally employs the training subspace for projection purposes, aligning mathematically with the principles of EC. This method assumes that the system varies smoothly with adjustable parameters, an idealization that does not always hold true in practice.
A notable example of this occurs around the dripline in nuclear physics, where systems transitioning across the threshold become open quantum systems, frequently exhibiting discontinuities. This phenomenon is known as the Wigner cusp~\cite{Wigner1948a,Michel2007a,Okolowicz2023a} and has been extensively studied within the field~\cite{Elhatisari2016a,Dassie2022a}. The presence of such discontinuities, particularly those emerging from asymptotic behavior~\cite{Wigner1948a,Michel2007a,Okolowicz2023a}, poses significant challenges for the traditional Galerkin projection method. These challenges necessitate a more adaptable RBM approach, one that can be possibly extended to handle resonance problems and other non-linear behaviors inherent to these complex quantum systems.

This article introduces a novel RBM that projects the Hamiltonian onto momentum space, thereby providing an alternative emulation technique for studying few-body bound state problems within open quantum systems. This method distinguishes itself through its efficiency, accuracy, and the flexibility it offers in the selection of projecting bases, thereby enhancing our understanding of the physical systems under examination.
We utilize a high-fidelity Hamiltonian generated by the Gamow coupled-channel (GCC) method \cite{Wang2017a, Wang2018a}, which is a three-body model in Jacobi coordinates with the Berggren basis. The Berggren basis encompasses scattering, bound, and resonant states on the same footing, thus providing an ideal framework for addressing complex issues prevalent in open quantum systems, such as two-proton decay \cite{Wang2021a, Wang2019a,Zhou2022a,Zhou2024a} and non-exponential decay phenomena \cite{Wang2023a}. The implementation of this innovative RBM is expected to catalyze further research utilizing more detailed microscopic models and to deepen our insight into the intricate properties of open quantum systems.

%%%%%%%%%%%%%%%%%%%%%%%%%%%%%%%%%%%%%%%%%%%%%%%%%%%%%%%%%%%%%%%%%%%%%%%%%%%%%%%%%%%%%%%%%%%%%%
%%%%%%%%%%%%%%%%%%%%%%%%%%%%%%%%%%%%%%%%% METHODS %%%%%%%%%%%%%%%%%%%%%%%%%%%%%%%%%%%%%%%%%%%%
%%%%%%%%%%%%%%%%%%%%%%%%%%%%%%%%%%%%%%%%%%%%%%%%%%%%%%%%%%%%%%%%%%%%%%%%%%%%%%%%%%%%%%%%%%%%%%
\section{Methods}\label{sec:methods}

%%%%%%%%%%%%%%%%%%%%%%%%%%% Gamow coupled-channel %%%%%%%%%%%%%%%%%%%%%%%%%%%
\subsection{Gamow coupled-channel approach}
As a starting point, an artificial $^6\text{Be}$ system, a frozen $\alpha$ core with two valence protons, is built to study the three-body emulation problems, and the Hamiltonian can be written as 
\begin{equation}\label{eq:Hamiltonian_6Be}
    \boldsymbol{H}(c) = \sum_i \frac{\boldsymbol{p}_i^2}{2m_i}+c \sum_{i>j} \boldsymbol{V}_{ij}(\boldsymbol{r}_{ij}) - \boldsymbol{T}_{\text{c.m.}},
\end{equation}
where ${\boldsymbol{p}_i^2}/{2m_i}$ is the kinetic energy for the $i$th cluster. Here, $\boldsymbol{V}_{ij}(\boldsymbol{r}_{ij})$ denotes the inter-cluster potential, which is modulated by a strength parameter $c$, reflecting the depth or intensity of the potential. The variable $\boldsymbol{r}_{ij}$ indicates the spatial separation between the $i$-th and $j$-th clusters. Additionally, the term $\boldsymbol{T}_{\text{c.m.}}$ effectively removes the kinetic energy associated with the center-of-mass motion.

To properly describe the asymptotic behavior, the three-body system's wave function is expressed using Jacobi coordinates with hyperspherical-harmonic oscillator basis~\cite{Wang2017a, Wang2021a}, thereby splitting the coordinates into six degrees of freedom with $\rho$ and $\Omega_5$ symbolizing the radial and angular components, respectively~\cite{Descouvemont2003a, De1983a, Kievsky2008a}. Consequently, the total wave function is written as~\cite{Yang2023a}
\begin{equation}\label{eq:GCC_wavefunction}
    \Psi^{J\pi}=\rho^{-5/2}\sum_{\gamma K}\int{C_{\gamma K}(k)\mathcal{B}^{\gamma K}(k,\rho)\mathcal{Y}_{\gamma K}^{JM}(\Omega_5)\mathrm{d}k},
\end{equation}
where $\mathcal{Y}_{\gamma K}^{JM}(\Omega_5)$ symbolizes the hyperspherical harmonics describing the hyper-angular dynamics. Here, $K$ denotes the hyperspherical quantum number, and $\gamma=\{S_{12}, S, \mathcal{l}_x, \mathcal{l}_y, L\}$ signifies a set of quantum numbers encompassing spin and orbital angular momenta configurations \cite{Wang2017a}.

In treating the hyper-radial dynamics, the Berggren ensemble $\mathcal{B}^{\gamma K}(k,\rho)$ is employed, which incorporates scattering states, bound states, and resonances on the same footing. This is achieved through an expansion of the wave function in the complex plane, providing a robust mathematical structure that facilitates the handling of complex resonant behaviors \cite{Bengtsson2013a, Michel2021a}. The total wave function, $\Psi^{J\pi}$, is then obtained by diagonalizing the Hamiltonian, characterized as a complex symmetric matrix within this theoretical framework. This methodological approach ensures a comprehensive and precise description of the system's quantum states under various conditions.

%%%%%%%%%%%%%%%%%%%%%%%%%%% Reduced Basis Method %%%%%%%%%%%%%%%%%%%%%%%%%%%
\subsection{Reduced basis method}

To obtain a fine-tuned Hamiltonian, it often becomes necessary to perform a series of iterative calculations that are similar in nature but vary the parameters within the interactions. This process not only necessitates redundant computational efforts but also demands substantial computational resources. In addressing these parametric intricacies, RBMs have proven to be highly effective.

Within the framework of RBMs, consider a scenario where a set of $N_b$ high-fidelity energy eigenvalues $E_n$, along with corresponding wave functions $\boldsymbol{\Psi}^{\rm RB}_n$ for $n=0,~1,~...,~N_b-1$, have been computed using the GCC method for Hamiltonian $\boldsymbol{H}(c_n)$. Electing these $N_b$ wave functions $\boldsymbol{\Psi}^{\rm RB}_n$ as the reduced basis, the solution $\boldsymbol{\Psi}_{\odot}$ for a different Hamiltonian $\boldsymbol{H}_{\odot}$, associated with a specific target parameter $c_{\odot}$, can be approximated by a linear combination:
\begin{equation}\label{eq:RBM_wavefunction_expansion}
    \boldsymbol{\Psi}_{\rm \odot} \approx \sum_{n=0}^{N_b-1}a_n\boldsymbol{\Psi}^{\rm RB}_n = \boldsymbol{X}_{\rm train}\cdot\boldsymbol{a}, 
\end{equation}
with $\boldsymbol{a}=[a_0,~a_1,~...,~a_{N_b-1}]^T$ representing the expansion coefficients within the training subspace $\boldsymbol{X}_{\rm train}$. It is imperative that $\boldsymbol{H}(c)$ exhibits smooth variability with $c$ to maintain the fidelity of RBM approximations~\cite{Frame2018a, Quarteroni2016a}.

The eigenvalue problem
\begin{equation}\label{eq:target_schrodingers_equation}
    \boldsymbol{H}_{\rm \odot}\boldsymbol{\Psi}_{\rm \odot}=E_{\rm \odot}\boldsymbol{\Psi}_{\rm \odot}
\end{equation}
subsequently transforms to the equation
\begin{equation}\label{eq:RBM_equation_to_be_solved_in_full_space}
    \boldsymbol{F}_{\rm \odot}(\boldsymbol{\Psi}_{\rm \odot}) = (\boldsymbol{H}_{\rm \odot}-E_{\rm \odot})\boldsymbol{\Psi}_{\rm \odot}=\boldsymbol{0}.
\end{equation}

With the Galerkin method~\cite{Melendez2022a,Bonilla2022a,Quarteroni2016a}, the projecting basis can usually be chosen as those training states or arbitrary function $\boldsymbol{\psi}_m$. As such, Equation~\eqref{eq:RBM_equation_to_be_solved_in_full_space} can be written as
\begin{equation}\label{eq:RBM_projection}
    \boldsymbol{\psi}_m^T\cdot\boldsymbol{F}_{\rm \odot}(\boldsymbol{X}_{\rm train}\cdot\boldsymbol{a})=0,
\end{equation}
where $m=0,~1,~...,~M_b-1$, and the solution for $\boldsymbol{a}$ is attainable provided $M_b \le N_b$. In the end, if $\boldsymbol{H}(c)$ changes smoothly, there would be a chance to get reasonable solutions with a proper choice of training and projecting subspace.

The general examples~\cite{Bonilla2022a, Sarkar2022a, Odell2024a} of RBM are similar to EC, whose projecting subspace is equivalent to the training one. However, when the system is unbound, the regular variation principle does not apply to the resonances, and there is a discontinuity between the bound and scattering spaces due to the threshold effect. Consequently, the standard EC method inadequately addresses resonance behavior. We respond by proposing a more versatile RBM, utilizing free-particle wave functions for the projection basis. While primarily illustrating bound-state solutions, our method lays the groundwork for resonance analysis and is benchmarked against classical RBM for performance.

Setting the projecting basis count to $N_b$, analogous to the training set, the designated subspace $\boldsymbol{Y}_{\rm proj}=[\boldsymbol{\psi}_{0},~\boldsymbol{\psi}_{1},~...,~\boldsymbol{\psi}_{N_b-1}]$ is composed of free-particle wave functions $\boldsymbol{\psi}_n = | k_n \gamma K \rangle$, where each $k_n$ is a momentum value randomly selected from the interval $[k^\text{train}_\text{min},~k^\text{train}_\text{max}]$. This selection strategy reflects the distinction between high-momentum subspaces, which correspond to the inner structure details in the coordinate space, and low-momentum subspaces, which are indicative of the asymptotic region. Such differentiation allows for the tailored extraction of effective nuclear information by filtering momentum intervals selectively.

\begin{figure}[!htb]
    \includegraphics[width = 1.0\hsize]{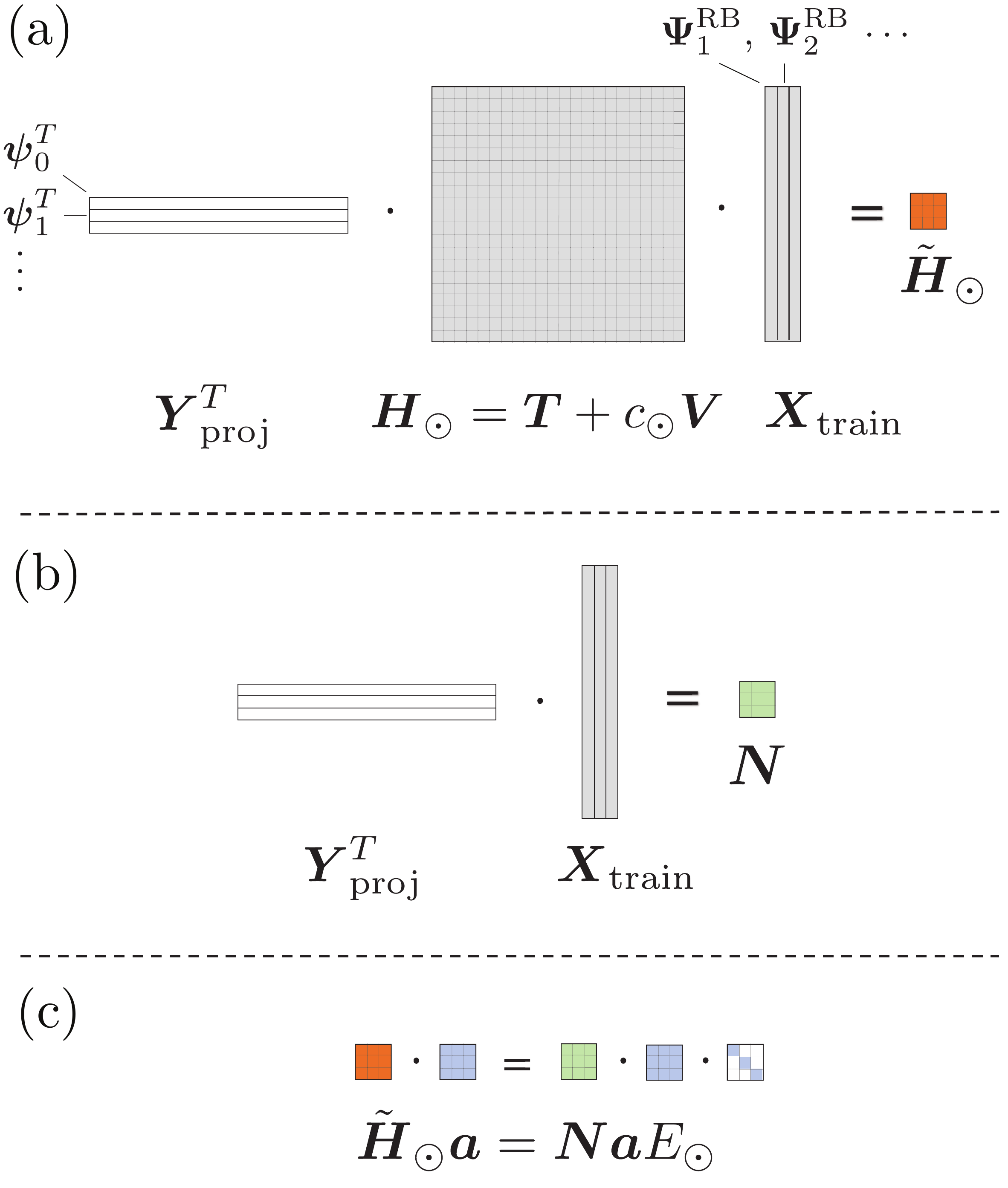}
    \caption{Projection process of RBM in the current framework. By selecting free-particle wave functions $\boldsymbol{\psi}$ in momentum space as the projection basis for training wave functions, this basis acts as a filter, transforming the Hamiltonian matrix $\boldsymbol{H}_\odot$ into a reduced-dimensional form $\Tilde{\boldsymbol{H}}_\odot$, illustrated in panel (a). Concurrently, the norm matrix $\boldsymbol{N}$, shown in panel (b), is constructed from overlaps between the projection and training wave function matrices. Finally, the target system's physical energy is obtained by diagonalizing the matrix $\boldsymbol{N}^{-1}\cdot\Tilde{\boldsymbol{H}}_\odot$, as shown in panel (c), according to Eq.~\eqref{eq:method-projc}. }
    \label{fig:projecting_operation_scheme}
\end{figure}

According to Eqs.~\eqref{eq:RBM_wavefunction_expansion} and ~\eqref{eq:RBM_projection}, the Schr$\ddot{\text{o}}$dinger equation can be reduced as:
\begin{align}{\label{eq:method-projc}}
    % \boldsymbol{Y}_{\rm proj}^T(\boldsymbol{H}_{\rm \odot}-E_{\rm \odot})\boldsymbol{X}_{\rm train}\cdot\boldsymbol{a} &= \boldsymbol{0}, \nonumber \\
    % \boldsymbol{Y}_{\rm proj}^T(\boldsymbol{H}_0+c_\odot\boldsymbol{H}_1)\boldsymbol{X}_{\rm training}\cdot\boldsymbol{a} &= E_{\rm \odot}\boldsymbol{Y}_{\rm proj}^T\boldsymbol{X}_{\rm training}\cdot\boldsymbol{a} \nonumber \\
    \boldsymbol{Y}_{\rm proj}^T\cdot\boldsymbol{F}_{\rm \odot}(\boldsymbol{X}_{\rm train}\cdot\boldsymbol{a}) &= \boldsymbol{0}, \nonumber \\
    \Tilde{\boldsymbol{H}}_\odot\boldsymbol{a} &= \boldsymbol{N}\boldsymbol{a}E_{\rm \odot}.
\end{align}
Here, $\Tilde{\boldsymbol{H}}_\odot = \boldsymbol{Y}_{\rm proj}^T\boldsymbol{H}_{\rm \odot}\boldsymbol{X}_{\rm train}$ denotes the projected target Hamiltonian, and the norm matrix $\boldsymbol{N} = \boldsymbol{Y}_{\rm proj}^T\boldsymbol{X}_{\rm train}$ reflects the projection of the training basis. Notably, employing the free-particle wave function in the complex momentum space as the Berggren basis facilitates the truncation of specific elements from the full Hamiltonian matrix, thus forming a submatrix that acts as a training filter. This process is illustrated in Fig.\ref{fig:projecting_operation_scheme}, where the Hamiltonian is transformed into a squared matrix configuration, thereby enabling the extraction of eigenvalues through diagonalization within a substantially reduced dimensional space, which enhances computational efficiency significantly.

%%%%%%%%%%%%%%%%%%%%%%%%%%% Model Space and Parameters %%%%%%%%%%%%%%%%%%%%%%%%%%%
\subsection{Model space and parameters}

Within this study, the phenomenological Woods-Saxon potential~\cite{Schwierz2007a}, with the spin-orbit term and Coulomb force included, is adopted to describe the interactions between the core and valence nucleons. The tunable parameters of the core-valence interaction for $^6$Be are chosen as $a = 0.7~\text{fm},~V_0 = 49.6~\text{MeV},~R_0 = 2.0~\text{fm},~V_\text{s.o.} = 39.5~\text{MeV}$, and $R_\text{s.o.} = 2.1~\text{fm}$, where the meanings of them can be found in Ref.~\cite{Wang2017a}. Specifically, $R_0$ and $R_\text{s.o.}$ are the radius of the central term and spin-orbit term, respectively. The interaction between the two valence protons is described by the finite-range Minnesota force with the same parameters as Ref.~\cite{Thompson1977a}, which forms the potential term of $\boldsymbol{H}$ together with the core-valence one. To form a deeply bound nucleus, we adjust the potential strength $c$ among the limitation $[1.5, 3.5]$ both for training and target points.

In the GCC model, the configurations are defined within Jacobi coordinates and characterized by quantum numbers $(K, \ell_x, \ell_y)$. Here, $\ell_x$ represents the orbital angular momentum of the proton pair relative to their center of mass, while $\ell_y$ denotes the orbital angular momentum of this pair with respect to the $\alpha$ core. The Pauli-forbidden states occupied by the core nucleons are eliminated according to Ref.\,\cite{Sparenberg1997a}. The computational framework restricts the model space such that $\max(\ell_x, \ell_y) \leq 8$, and the maximum hyperspherical quantum number is set to $K_{\max} = 16$ \cite{Wang2017a}. To incorporate the effects of the continuum, the Berggren basis is employed for channels where $K_{\max} \leq 3$, supplemented by a Harmonic Oscillator (HO) basis with oscillator length $b = 1.75$ fm and $N_{\max} = 40$ for channels involving higher angular momenta. The complex-momentum contour utilized in the Berggren basis is defined by $\tilde{k} = 0 \rightarrow 0.2-0.05i \rightarrow 0.3 \rightarrow 0.4 \rightarrow 0.5 \rightarrow 0.8 \rightarrow 1.2 \rightarrow 2 \rightarrow 4 \rightarrow 6$ fm$^{-1}$, with each segment containing 120 discretized scattering states.

In the realm of RBM, different intervals within the momentum space are selected to serve as the filter or projection space, facilitating the assessment of our computational emulator's efficiency.

%%%%%%%%%%%%%%%%%%%%%%%%%%%%%%%%%%%%%%%%%%%%%%%%%%%%%%%%%%%%%%%%%%%%%%%%%%%%%%%%%%%%%%%%%%%%%%
%%%%%%%%%%%%%%%%%%%%%%%%%%%%%%%%%% RESULTS AND DISCUSSIONS %%%%%%%%%%%%%%%%%%%%%%%%%%%%%%%%%%%
%%%%%%%%%%%%%%%%%%%%%%%%%%%%%%%%%%%%%%%%%%%%%%%%%%%%%%%%%%%%%%%%%%%%%%%%%%%%%%%%%%%%%%%%%%%%%%

\section{Results and Discussions}

%%%%%%%%%%%%%%%%%%%%%%%%%%% Three-body wave functions and the PCA %%%%%%%%%%%%%%%%%%%%%%%%%%%
\subsection{Three-body wave functions and the principal component analysis}\label{sec:three_body_wavefunctions_and_PCA}

\begin{figure}[!htb]
    \includegraphics[width = 0.95\hsize]{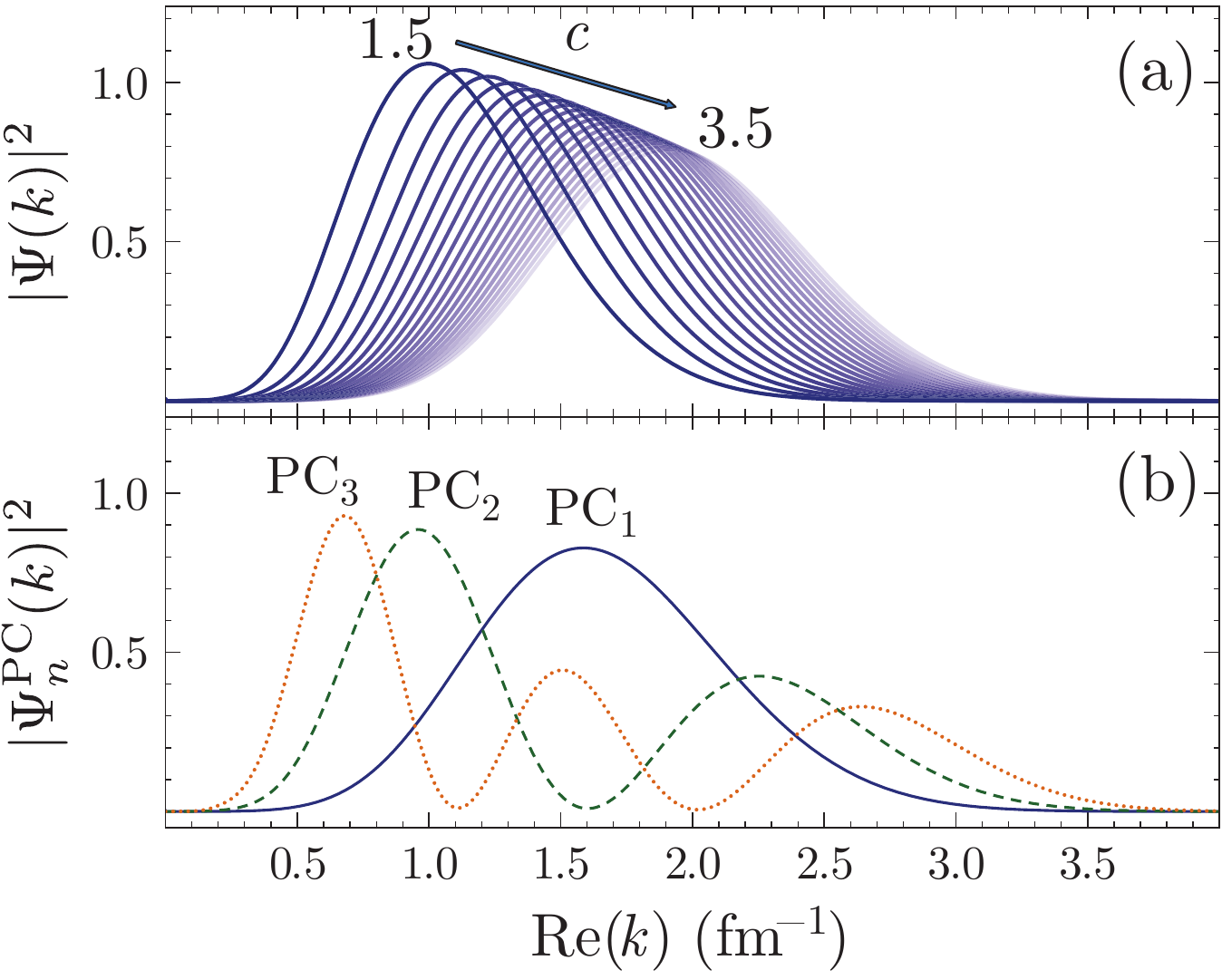}
    \caption{(a) Squared norm of the wavefunctions as a function of the strength of the single-particle potential $c$ within the range of $1.5$ to $3.5$. (b) Associated principal components (PCs).}
    \label{fig:training_wavefunctions}
\end{figure}

Using the Hamiltonian mentioned in Sec.~\ref{sec:methods}, the ground state energy of $\rm ^{6}Be$ as a function of the potential strength, $c$, is calculated. As $c$ increases, the system trends towards a bound state due to the deepening of the potential well~\cite{Dassie2022a}. Consequently, Figure~\ref{fig:training_wavefunctions}a shows how the three-body wave functions of the bound states change with the strength of the potential in the momentum space. They exhibit significant diffusion, which implies a localized distribution in the coordinate space due to the bound nature, consistent with Heisenberg's uncertainty principle~\cite{Sakurai2020a}. In particular, the wave functions smoothly transition with varying $c$, demonstrating their suitability for assembling a training subspace for emulation purposes. 

\begin{figure}[!htb]
    \includegraphics[width = 0.95\hsize]{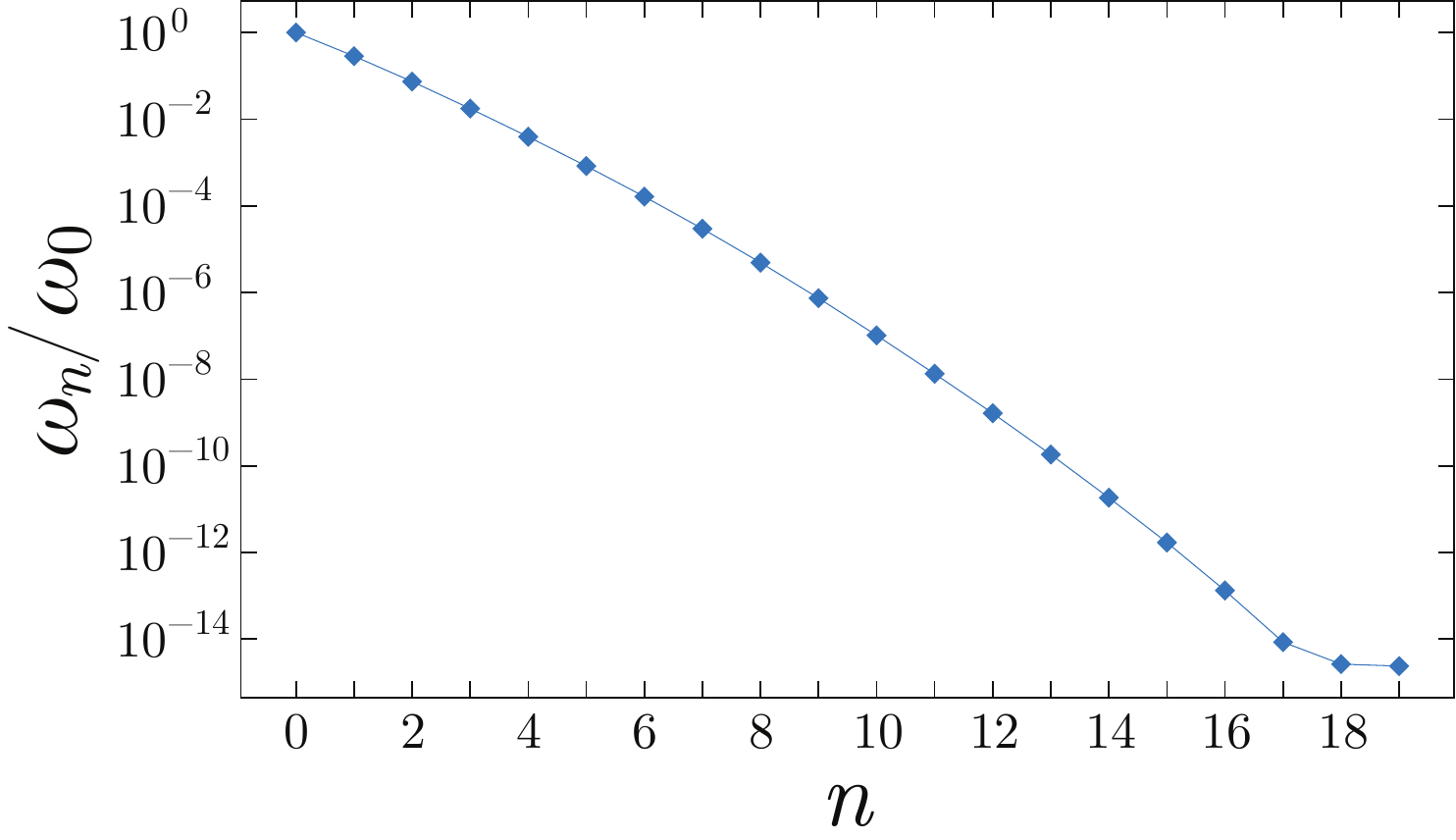}
    \caption{Relative weights of PCs associated with the wave functions in the training subspace. $\omega_0$ represents the weight of the first principal component, while $\omega_n$ signifies the weight of the $n$-th principal component.}\label{fig:training_wavefunctions_singular_values}
\end{figure}

Usually, most RBMs will adopt Principal Component Analysis (PCA) as the efficiency diagnosis. An ideal scenario is achieved when the weights of PCs in the training basis decay exponentially, indicating a problem well-suited for emulation and a training space conducive to the expansion of target wave functions~\cite{Quarteroni2016a, Quarteroni2011a}. By selecting PC weights above a certain threshold number, we can tailor the desired accuracy and mitigate ill-conditioned issues. Figure~\ref{fig:training_wavefunctions}b presents the PCA-transformed wave functions, with the dark blue line representing the most significant component. This component encapsulates the quintessential physical information about the bound states, serving as the median of all wave functions ranging from $c=1.5$ to $c=3.5$. The green dashed line and the orange dot line illustrate the shape of the second and third important components, respectively, with singular values as low as one magnitude than the first component, which can be found in Fig.~\ref{fig:training_wavefunctions_singular_values}. Most of the physics information can be obtained from these three PCA components. 

% At high momenta more than 3 fm$^{-1}$, near the nucleus' core, the repulsive nature of nuclear forces becomes evident, leading to a nearly zero density distribution~\cite{}. In contrast, at very low momenta — within the asymptotic range — the instability of the nucleus results in wave function undulations in the coordinate space~\cite{}. 
% In fact, the wave number $k$ has an influence on the wave function in the coordinate space. At high momenta, $k \geq 3$ fm$^{-1}$, for example, the wave function changes faster with the coordinates, while it is lower at very low momenta. Thus, the wave functions at the asymptotic region are orthogonal between higher momenta and lower ones.
% Accordingly, we have designated the inner components as training filters to capture essential nuclear structure insights. This choice also supports the motivation for selecting free-particle wave functions as the projecting basis.

The analysis of wave functions and PC behavior reveals that in momentum space, the wave functions' low- and high-momentum components evolve more gradually than their medium-momentum counterparts. Given that momentum $k$ correlates with the wave function's frequency in the asymptotic region, it is feasible to utilize free-particle waves, represented by specific momenta, as a projecting basis. This basis acts effectively as a high-frequency filter, orthogonal to low-lying resonances characterized by relatively low momenta, to discern the intrinsic structural information of the atomic nucleus. For example, a projecting interval at relatively high momentum correlates with the nucleus's short-range behavior, and conversely, low momentum correlates with long-range characteristics. Selecting a high-frequency filter, however, tends to overlook the details of the asymptotic wave function, concentrating instead on the internal structure of the system. It is notable that the internal wave function typically exhibits smooth transitions from bound states to even resonances, making it an ideal projection subspace for RBM. By integrating PCA, it is possible to further refine the training subspace, thereby preserving essential physical information without significant loss.

% In the GCC model, the original wave function, as extracted from the Hamiltonian, bears two distinct properties. First, as discussed in Sec. [section reference needed], the wave function is discretized with specific weights at each momentum point, necessitating a transformation before one can procure physical plots, akin to Fig. a). 
% Secondly, the wave function is a direct sum of the radial one of each configuration. There are some 'black box' components when PCA is used to extract information from the training space. Fig. illustrates that the 'original wave functions' are capable of decomposition because the singular values are decaying exponentially. 

%%%%%%%%%%%%%%%%%%%%%%%%%%% Computational performance of RBM %%%%%%%%%%%%%%%%%%%%%%%%%%%
\subsection{Computational performance of RBM}

The effectiveness of the RBM is contingent upon the proper definition of the spaces for training and projection. To evaluate the computational efficacy of our approach, we select training filters as random, ensuring that the real component of the complex momenta, Re$(k)$, lies within the range $[0.5, 2.5]$ fm$^{-1}$. The performance of the RBM is then assessed by sampling 100 target points where the parameter $c$ varies from $1.5$ to $3.5$, hereafter referred to as $c_\odot$. We compare the eigenvalues, which represent the ground-state energies, with the precise solutions obtained through the GCC model within a complete space of dimension 3900. The relative error at each target point is quantified as follows:
\begin{equation}
    \epsilon(c_\odot)=\log_{10}\left|\frac{E(c_\odot)-E_{\rm exact}(c_\odot)}{E_{\rm exact}(c_\odot)}\right|,
\end{equation}
where $E(c_\odot)$ represents the energy estimations derived from the methodology under test.

\begin{figure}
    \includegraphics[width = 0.9\hsize]{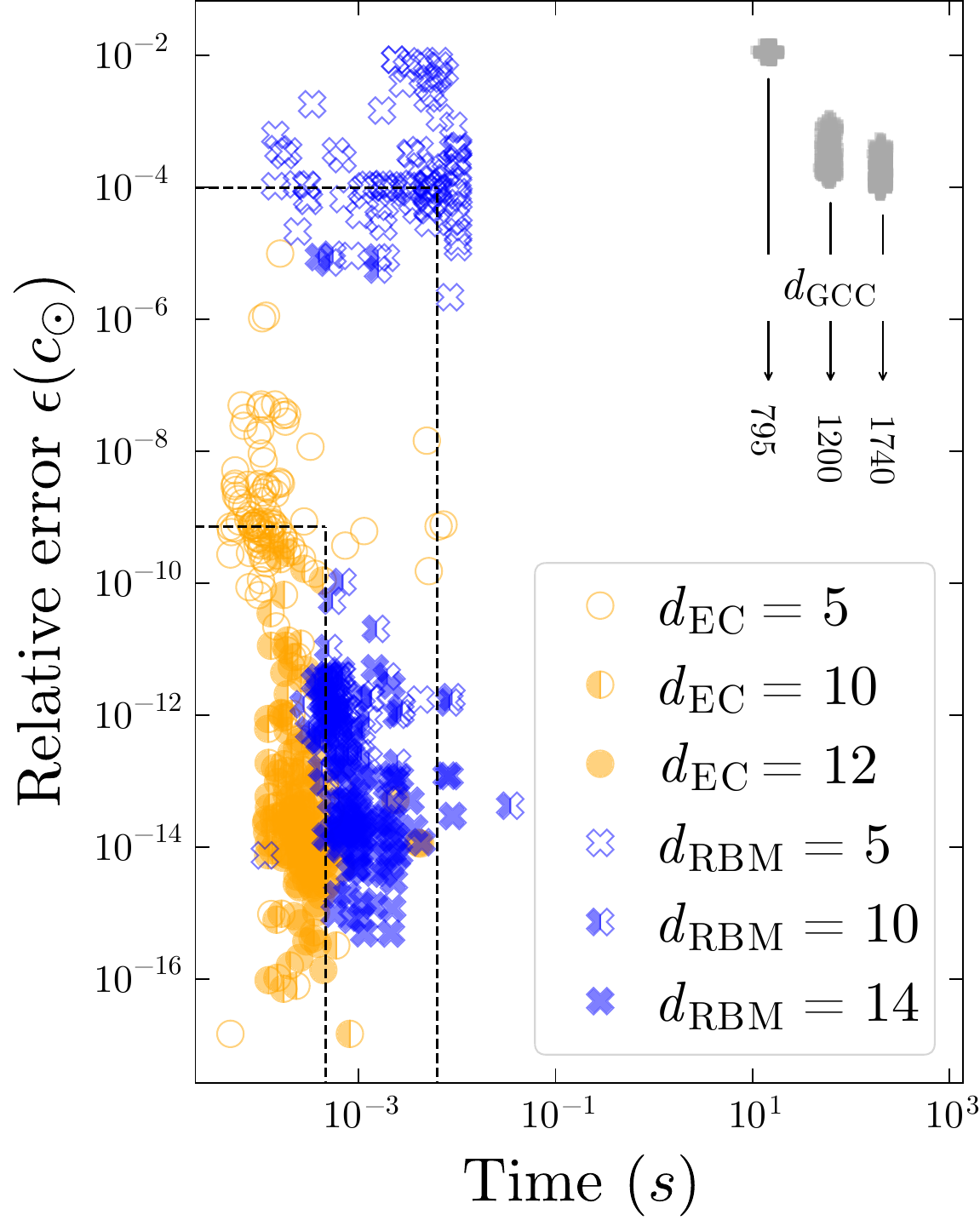}
    \caption{Relative errors $\epsilon$ and computational consumption of RBM compared with the EC and GCC methods.  A total of 100 target points, each characterized by varying the target potential strengths $c_{\odot}$, were selected for this evaluation. 
    % The blue crosses with zero-, half-, and full-filled are calculated by our new RBM with the subspace size of 5, 10, and 14, respectively. The orange dots with zero-, half-, and full-filled are calculated by EC, the general RBM, with the subspace size of 5, 10, and 12, respectively. Besides, the green crosses are calculated by GCC with different sizes of full space in 795, 1200, and 1740. The horizontal axis represents the time spent for solving the Schr$\ddot{\text{o}}$dinger equation of each point, and the vertical axis is the relative error of each point. It is obvious that the emulators have an advantage in performing large repeat calculations quickly and accurately compared to its microscopic model.
    }
    \label{fig:error_time}
\end{figure}

As illustrated in Fig.~\ref{fig:error_time}, the predictions of our RBM demonstrate that over 90\% of the evaluated points exhibit a relative error, $\epsilon(c_\odot)$, of less than $10^{-4}$. Notably, a significant clustering of data points around $\epsilon(c_\odot) \approx 10^{-4}$, marked by hollow blue crosses, corresponds to a training subspace dimension of only five. The dimensionality of the emulators, denoted as $d_{\text{RBM}}$ or $d_{\text{EC}}$, is influenced by both the precision of the PCA weights and the quantity of training wave functions involved. For example, $d_{\text{RBM}}$ takes values such as $5, 10, 14$, corresponding to the settings of $5, 10, 20$ training wave functions, respectively, while maintaining the precision of singular values at a constant $10^{-12}$ to eliminate trivial components of the wave function.

\begin{figure}[!htb]
    \includegraphics[width = 0.9\hsize]{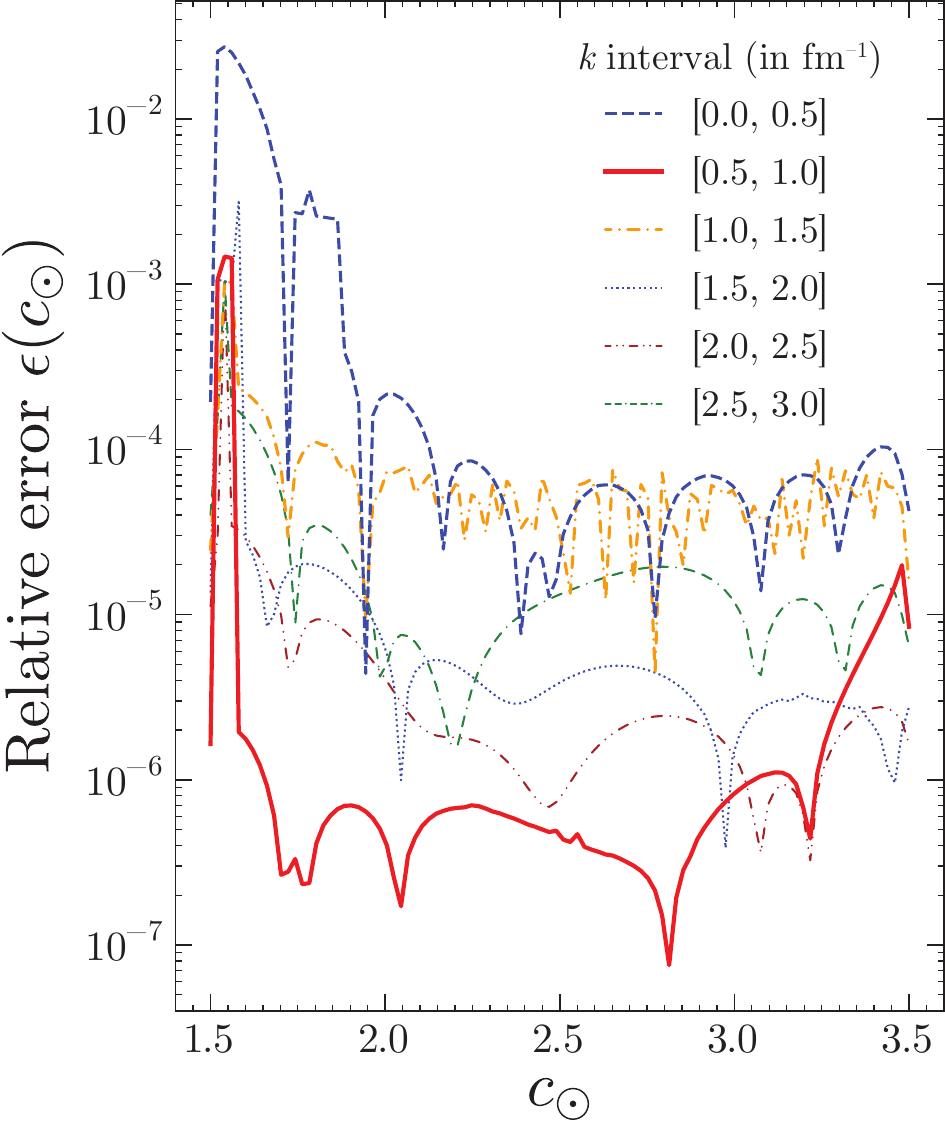}
    \caption{Relative error $\epsilon$ of each target point $c_\odot$ with different training filters. The red line holds the most accurate calculation of the RBM, while the momemtum $k$ of the training filters is varied within the interval of 0.5 to 1.0 fm$^{-1}$.}
    \label{fig:different_training_masks}
\end{figure}

The training basis of RBM, as delineated in Eq.~\eqref{eq:RBM_wavefunction_expansion}, thus correlates with the number of principal components rather than with the raw count of wave functions. This correlation results in $d_{\text{RBM}}$ stabilizing at 14, regardless of increases in the number of training wave functions to $30, 40,$ or $50$. Such stabilization effectively captures the critical information within the training subspace without incurring increased errors. This behavior aligns with the detailed analysis provided in Sec.~\ref{sec:three_body_wavefunctions_and_PCA}, underscoring the efficacy of the RBM in managing dimensionality while retaining computational accuracy.

Time efficiency is also a critical metric in computing the eigenvalues for each target point. As shown in Fig.~\ref{fig:error_time}, the time required to solve the Schr{\"{o}}dinger equation using emulators is significantly less compared to the GCC method. Specifically, the emulator solution time is approximately $\SI{e-3}{\second}$ per point, whereas GCC requires a duration that is four orders of magnitude longer. Additionally, the error predicted by the RBM decreases markedly to a value substantially lower than that observed with GCC by only slightly increasing the dimension of the training subspace. This reduced computational time of the emulators, relative to GCC, is attributed to their streamlined dimensionality and efficient error minimization. Consequently, our RBM implementation not only substantially reduces computational resource demands but also achieves performance nearly equivalent to EC for three-body bound states. This enhancement underscores the effectiveness of RBM in delivering precise and efficient solutions in quantum mechanical simulations.

Figure~\ref{fig:different_training_masks} explores the variability in the behavior of filter or projection spaces, shedding light on how the error for each sampled point with adjustable parameter $c$ varies with the training filters defined across different momentum space intervals. Notably, the red solid line, which represents masks with Re$(k_n)$ in the interval $[0.5,\ 1.0]\ \text{fm}^{-1}$, delivers the most accurate results. This momentum range approximately corresponds to energies between $[5,\ 20]\ \text{MeV}$, a crucial energy spectrum for lower-energy nuclear physics research that is rich with both structural and continuum data. As the energy increases, the relevance of structural details diminishes, while the scattering aspects in reactions become more pronounced. Consequently, errors in simulations using filters with higher Re$(k_n)$ values, as represented by the dotted lines, increase by approximately an order of magnitude. On the other hand, the selection of excessively low Re$(k_n)$ values, depicted by the blue dashed line in Fig.~\ref{fig:different_training_masks}, proves to be less optimal for emulation purposes due to their greater distance from the nucleus's core and the significant influence of asymptotic behaviors on the state. Therefore, the strategic choice of training filters in RBM not only preserves the computational efficiency of the emulator but also enriches our understanding of the structural characteristics of nuclei.

%%%%%%%%%%%%%%%%%%%%%%%%%%%%%%%%%%%%%%%%%%%%%%%%%%%%%%%%%%%%%%%%%%%%%%%%%%%%%%%%%%%%%%%%%%%%%%
%%%%%%%%%%%%%%%%%%%%%%%%%%%%%%%%%%%%%%%% SUMMARY %%%%%%%%%%%%%%%%%%%%%%%%%%%%%%%%%%%%%%%%%%%%%
%%%%%%%%%%%%%%%%%%%%%%%%%%%%%%%%%%%%%%%%%%%%%%%%%%%%%%%%%%%%%%%%%%%%%%%%%%%%%%%%%%%%%%%%%%%%%%

\section{Summary}

In this study, we introduce a novel adaptation of the Reduced Basis Method tailored for nuclear physics, which employs free-particle wavefunctions as the projecting basis. By integrating this with the three-body Gamow coupled-channel model, which serves as the high-fidelity method, we have developed an emulator to analyze the bound-state energy of an artificial $^6$Be system. This approach is proven to be both rapid and precise for solving the parametric Schr{\"{o}}dinger equation by modulating the potential strength. Moreover, our methodology facilitates the extraction of crucial physical insights within the subspace by carefully selecting the momentum interval of the training filters. Details of these analyses are comprehensively presented in this article. Owing to its flexibility in choosing the projecting subspace, this method holds significant potential for extension to studies in nuclear open quantum systems, particularly in examining resonances. Such flexibility is instrumental in pinpointing critical resonance information, enhancing our understanding of these complex quantum systems.

%%%%%%%%%%%%%%%%%%%%%%%%%%%%%%%%%%%%%%%%%%%%%%%%%%%%%%%%%%%%%%%%%%%%%%%%%%%%%%%%%%%%%%%%%%%%%%
%%%%%%%%%%%%%%%%%%%%%%%%%%%%%%%%%%%% ACKNOWLEDGEMENTS %%%%%%%%%%%%%%%%%%%%%%%%%%%%%%%%%%%%%%%%
%%%%%%%%%%%%%%%%%%%%%%%%%%%%%%%%%%%%%%%%%%%%%%%%%%%%%%%%%%%%%%%%%%%%%%%%%%%%%%%%%%%%%%%%%%%%%%

% \printbibliography
\bibliography{references.bib}

\begin{thebibliography}{10}

\bibitem{Carlson2015a}
J.~Carlson, S.~Gandolfi, F.~Pederiva, et~al.
\newblock Quantum {Monte Carlo} methods for nuclear physics.
\newblock {\em Rev. Mod. Phys.}, 87(3):1067, 2015.

\bibitem{Navratil2016a}
P.~Navr{\'a}til, S.~Quaglioni, G.~Hupin, et~al.
\newblock Unified ab initio approaches to nuclear structure and reactions.
\newblock {\em Phys. Scr.}, 91(5):053002, 2016.

\bibitem{Hergert2016a}
H.~Hergert, S.K. Bogner, T.D. Morris, et~al.
\newblock The in-medium similarity renormalization group: A novel ab initio method for nuclei.
\newblock {\em Phys. Rep.}, 621:165--222, 2016.

\bibitem{Michel2021a}
N.~Michel and M.~P{\l}oszajczak.
\newblock {\em Gamow Shell Model, The Unified Theory of Nuclear Structure and Reactions}.
\newblock Springer Cham, 1 edition, 2021.

\bibitem{Li2021a}
J.~Li, Y.~Ma, N.~Michel, et~al.
\newblock Recent progress in {Gamow} shell model calculations of drip line nuclei.
\newblock {\em Physics}, 3(4):977--997, 2021.

\bibitem{Michel2002a}
N.~Michel, W.~Nazarewicz, M.~P{\l}oszajczak, et~al.
\newblock Gamow shell model description of neutron-rich nuclei.
\newblock {\em Phys. Rev. Lett.}, 89(4):042502, 2002.

\bibitem{Boehnlein2022a}
A.~Boehnlein, M.~Diefenthaler, N.~Sato, et~al.
\newblock Colloquium: {Machine learning} in nuclear physics.
\newblock {\em Rev. Mod. Phys.}, 94(3):031003, 2022.

\bibitem{He2023b}
W.~He, Q.~Li, Y.~Ma, et~al.
\newblock Machine learning in nuclear physics at low and intermediate energies.
\newblock {\em Sci. China Phys., Mech. \& Astro.}, 66(8):282001, 2023.

\bibitem{Shang2022a}
T.~S. Shang, J.~Li, and Z.~M. Niu.
\newblock Prediction of nuclear charge density distribution with feedback neural network.
\newblock {\em Nucl. Sci. Tech}, 33:153, 2022.

\bibitem{Alhassan2022a}
E.~Alhassan, D.~Rochman, A.~Vasiliev, et~al.
\newblock Iterative {Bayesian Monte Carlo} for nuclear data evaluation.
\newblock {\em Nucl. Sci. Tech}, 33:50, 2022.

\bibitem{Brunton2019a}
S.~L. Brunton and J.~N. Kutz.
\newblock {\em Data-driven science and engineering: {Machine learning}, dynamical systems, and control}.
\newblock Cambridge University Press, 1 edition, 2019.

\bibitem{He2019a}
H.~He, S.~Jin, C.-K. Wen, et~al.
\newblock Model-driven deep learning for physical layer communications.
\newblock {\em IEEE Wireless Commun.}, 26(5):77--83, 2019.

\bibitem{Ma2023a}
Y.~G. Ma, L.~G. Pang, R.~Wang, et~al.
\newblock Phase transition study meets machine learning.
\newblock {\em Chin. Phys. Lett.}, 40:122101, 2023.

\bibitem{He2023a}
W.~B. He, Y.~G. Ma, L.~G. Pang, et~al.
\newblock High-energy nuclear physics meets machine learning.
\newblock {\em Nucl. Sci. Tech.}, 34(88):88, 2023.

\bibitem{Wang2020a}
R.~Wang, Y.~G. Ma, R.~Wada, et~al.
\newblock Nuclear liquid-gas phase transition with machine learning.
\newblock {\em Phys. Rev. Res.}, 2:043202, 2020.

\bibitem{He2021a}
J.~He, W.~B. He, Y.~G. Ma, et~al.
\newblock Machine-learning-based identification for initial clustering structure in relativistic heavy-ion collisions.
\newblock {\em Phys. Rev. C}, 104(4):044902, 2021.

\bibitem{Utama2016a}
R.~Utama, J.~Piekarewicz, and H.~B. Prosper.
\newblock Nuclear mass predictions for the crustal composition of neutron stars: A {Bayesian} neural network approach.
\newblock {\em Phys. Rev. C}, 93(1):014311, 2016.

\bibitem{Neufcourt2020a}
L.~Neufcourt, Y.~Cao, S.~Giuliani, et~al.
\newblock Beyond the proton drip line: {Bayesian} analysis of proton-emitting nuclei.
\newblock {\em Phys. Rev. C}, 101(1):014319, 2020.

\bibitem{Neufcourt2020b}
L.~Neufcourt, Y.~Cao, S.~A. Giuliani, et~al.
\newblock Quantified limits of the nuclear landscape.
\newblock {\em Phys. Rev. C}, 101(4):044307, 2020.

\bibitem{Huang2022a}
Y.~Huang, L.-G. Pang, X.~Luo, et~al.
\newblock Probing criticality with deep learning in relativistic heavy-ion collisions.
\newblock {\em Phys. Lett. B}, 827:137001, 2022.

\bibitem{Kejzlar2023a}
V.~Kejzlar, L.~Neufcourt, and W.~Nazarewicz.
\newblock Local {Bayesian Dirichlet} mixing of imperfect models.
\newblock {\em Sci. Rep.}, 13(1):19600, 2023.

\bibitem{Semposki2022a}
A.~C. Semposki, R.~J. Furnstahl, and D.~R. Phillips.
\newblock Interpolating between small- and large-$g$ expansions using {Bayesian} model mixing.
\newblock {\em Phys. Rev. C}, 106:044002, 2022.

\bibitem{Pang2023a}
L.~G. Pang and X.~N. Wang.
\newblock Bayesian analysis of nuclear equation of state at high baryon density.
\newblock {\em Nucl. Sci. Tech.}, 34(194):194, 2023.

\bibitem{Odell2024a}
D.~Odell, P.~Giuliani, K.~Beyer, et~al.
\newblock Rose: A reduced-order scattering emulator for optical models.
\newblock {\em Phys. Rev. C}, 109(4):044612, 2024.

\bibitem{Smith2024a}
A.~J. Smith, C.~Hebborn, F.~M. Nunes, et~al.
\newblock Uncertainty quantification in $(p,n)$ reactions, 2024.
\newblock {\color{blue}\href{https://arxiv.org/pdf/2403.18629}{arXiv:2403.18629}}.

\bibitem{Giuliani2023a}
P.~Giuliani, K.~Godbey, E.~Bonilla, et~al.
\newblock Bayes goes fast: {Uncertainty} quantification for a covariant energy density functional emulated by the reduced basis method, 2023.
\newblock {\color{blue}\href{https://www.frontiersin.org/articles/10.3389/fphy.2022.1054524}{Front. Phys. \bf{10}}}.

\bibitem{Giuliani2024a}
P.~Giuliani, K.~Godbey, V.~Kejzlar, and W.~Nazarewicz.
\newblock Model orthogonalization and bayesian forecast mixing via principal component analysis.
\newblock {\em Phys. Rev. Res.}, 6:033266, Sep 2024.

\bibitem{Bai2021a}
D.~Bai and Z.~Ren.
\newblock Generalizing the calculable {R-matrix} theory and eigenvector continuation to the incoming-wave boundary condition.
\newblock {\em Phys. Rev. C}, 103(1):014612, 2021.

\bibitem{ZhangX2024a}
X.~Zhang.
\newblock A non-hermitian quantum mechanics approach for extracting and emulating continuum physics based on bound-state-like calculations, 2024.
\newblock {\color{blue}\href{https://arxiv.org/abs/2408.03309}{arXiv:2408.03309}}.

\bibitem{Yapa2023a}
N.~Yapa, K.~Fossez, and S.~K{\"o}nig.
\newblock Eigenvector continuation for emulating and extrapolating two-body resonances.
\newblock {\em Phys. Rev. C}, 107(6):064316, 2023.

\bibitem{Yapa2024a}
N.~Yapa, S.~K{\"o}nig, and K.~Fossez.
\newblock Towards scalable bound-to-resonance extrapolations for few-and many-body systems, 2024.
\newblock {\color{blue}\href{https://arxiv.org/abs/2409.03116}{arXiv:2409.03116}}.

\bibitem{Frame2018a}
D.~Frame, R.~He, I.~Ipsen, et~al.
\newblock Eigenvector continuation with subspace learning.
\newblock {\em Phys. Rev. Lett.}, 121(3):032501, 2018.

\bibitem{Frame2019a}
D.~K. Frame.
\newblock {\em Ab initio simulations of light nuclear systems using eigenvector continuation and auxiliary field Monte Carlo}.
\newblock PhD thesis, Michigan State University, 2019.

\bibitem{Sarkar2022a}
A.~Sarkar and D.~Lee.
\newblock Self-learning emulators and eigenvector continuation.
\newblock {\em Phys. Rev. Res.}, 4(2):023214, 2022.

\bibitem{Sarkar2021a}
A.~Sarkar and D.~Lee.
\newblock Convergence of eigenvector continuation.
\newblock {\em Phys. Rev. Lett.}, 126(3):032501, 2021.

\bibitem{Melendez2022a}
J.~A. Melendez, C.~Drischler, R.~J. Furnstahl, et~al.
\newblock Model reduction methods for nuclear emulators.
\newblock {\em J. Phys. G: Nucl. Part. Phys.}, 49(10):102001, 2022.

\bibitem{Drischler2023a}
C.~Drischler, J.~A. Melendez, R.~J. Furnstahl, et~al.
\newblock {BUQEYE} guide to projection-based emulators in nuclear physics.
\newblock {\em Front. Phys.}, 10:1092931, 2023.

\bibitem{Quarteroni2016a}
A.~Quarteroni, A.~Manzoni, and F.~Negri.
\newblock {\em Reduced Basis Methods for Partial Differential Equations}.
\newblock Springer Cham, 1 edition, 2016.

\bibitem{Bonilla2022a}
E.~Bonilla, P.~Giuliani, K.~Godbey, et~al.
\newblock Training and projecting: A reduced basis method emulator for many-body physics.
\newblock {\em Phys. Rev. C}, 106(5):054322, 2022.

\bibitem{Duguet2024a}
T.~Duguet, A.~Ekstr{\"o}m, R.~J. Furnstahl, et~al.
\newblock Colloquium: {Eigenvector continuation} and projection-based emulators.
\newblock {\em Rev. of Mod. Phys.}, 96(3):031002, 2024.

\bibitem{Kutz2016a}
J.~N. Kutz, S.~L. Brunton, B.~W. Brunton, et~al.
\newblock {\em Dynamic mode decomposition: data-driven modeling of complex systems}.
\newblock SIAM, 1 edition, 2016.

\bibitem{Wigner1948a}
E.~P. Wigner.
\newblock On the behavior of cross sections near thresholds.
\newblock {\em Phys. Rev.}, 73(9):1002, 1948.

\bibitem{Michel2007a}
N.~Michel, W.~Nazarewicz, and M.~P{\l}oszajczak.
\newblock Threshold effects in multichannel coupling and spectroscopic factors in exotic nuclei.
\newblock {\em Phys. Rev. C}, 75(3):031301, 2007.

\bibitem{Okolowicz2023a}
J.~Oko{\l}owicz, M.~P{\l}oszajczak, and W.~Nazarewicz.
\newblock Near-threshold resonances in $^{11}\text{C}$ and the $^{10}\text{B}(p, \alpha)^7\text{Be}$ aneutronic reaction.
\newblock {\em Phys. Rev. C}, 107(2):L021305, 2023.

\bibitem{Elhatisari2016a}
S.~Elhatisari, N.~Li, A.~Rokash, et~al.
\newblock Nuclear binding near a quantum phase transition.
\newblock {\em Phys. Rev. Lett.}, 117(13):132501, 2016.

\bibitem{Dassie2022a}
A.~C. Dassie, F.~Gerdau, F.~J. Gonzalez, et~al.
\newblock Illustrations of loosely bound and resonant states in atomic nuclei.
\newblock {\em Am. J. Phys.}, 90(2):118--125, 2022.

\bibitem{Wang2017a}
S.~M. Wang, N.~Michel, W.~Nazarewicz, et~al.
\newblock Structure and decays of nuclear three-body systems: the {Gamow} coupled-channel method in {Jacobi} coordinates.
\newblock {\em Phys. Rev. C}, 96(4):044307, 2017.

\bibitem{Wang2018a}
S.~M. Wang and W.~Nazarewicz.
\newblock Puzzling two-proton decay of $^{67}\text{Kr}$.
\newblock {\em Phys. Rev. Lett.}, 120(21):212502, 2018.

\bibitem{Wang2021a}
S.~M. Wang and W.~Nazarewicz.
\newblock Fermion pair dynamics in open quantum systems.
\newblock {\em Phys. Rev. Lett.}, 126(14):142501, 2021.

\bibitem{Wang2019a}
S.~M. Wang, W.~Nazarewicz, R.~J. Charity, et~al.
\newblock Structure and decay of the extremely proton-rich nuclei $^{11, 12}\text{O}$.
\newblock {\em Phys. Rev. C}, 99(5):054302, 2019.

\bibitem{Zhou2022a}
L.~Zhou, S.~M. Wang, D.~Q. Fang, et~al.
\newblock Recent progress in two-proton radioactivity.
\newblock {\em Nucl. Sci. Tech.}, 33(8):105, 2022.

\bibitem{Zhou2024a}
L.~Zhou, D.~Q. Fang, S.~M. Wang, et~al.
\newblock Structure and 2p decay mechanism of $^{18}\mathrm{Mg}$.
\newblock {\em Nucl. Sci. Tech.}, 35(6):107, 2024.

\bibitem{Wang2023a}
S.~M. Wang, W.~Nazarewicz, A.~Volya, et~al.
\newblock Probing the nonexponential decay regime in open quantum systems.
\newblock {\em Phys. Rev. Res.}, 5(2):023183, 2023.

\bibitem{Descouvemont2003a}
P.~Descouvemont, C.~Daniel, and D.~Baye.
\newblock Three-body systems with {Lagrange-mesh} techniques in hyperspherical coordinates.
\newblock {\em Phys. Rev. C}, 67:044309, 2003.

\bibitem{De1983a}
M.~F. De~la Ripelle.
\newblock The potential harmonic expansion method.
\newblock {\em Ann. Phys.}, 147(2):281--320, 1983.

\bibitem{Kievsky2008a}
A.~Kievsky, S.~Rosati, M.~Viviani, et~al.
\newblock A high-precision variational approach to three-and four-nucleon bound and zero-energy scattering states.
\newblock {\em J. Phys. G: Nucl. Part. Phys.}, 35(6):063101, 2008.

\bibitem{Yang2023a}
Y.~H. Yang, Y.~G. Ma, S.~M. Wang, et~al.
\newblock Structure and decay mechanism of the low-lying states in $^{9}\mathrm{Be}$ and $^{9}\mathrm{B}$.
\newblock {\em Phys. Rev. C}, 108:044307, 2023.

\bibitem{Bengtsson2013a}
J.~Bengtsson, P.~Granstr{\"o}m, O.~Embr{\'e}us, et~al.
\newblock {\em Quantum Resonances in a Complex-Momentum Basis}.
\newblock PhD thesis, Chalmers University of Technology, 2013.

\bibitem{Schwierz2007a}
N.~Schwierz, I.~Wiedenhover, and A.~Volya.
\newblock Parameterization of the {Woods-Saxon} potential for shell-model calculations, 2007.
\newblock {\color{blue}\href{https://arxiv.org/abs/0709.3525}{arXiv:0709.3525}}.

\bibitem{Thompson1977a}
D.~R. Thompson, M.~LeMere, and Y.~C. Tang.
\newblock Systematic investigation of scattering problems with the resonating-group method.
\newblock {\em Nucl. Phys.}, 286(1):53--66, 1977.

\bibitem{Sparenberg1997a}
J.-M. Sparenberg and D.~Baye.
\newblock Supersymmetry between phase-equivalent coupled-channel potentials.
\newblock {\em Phys. Rev. Lett.}, 79(20):3802, 1997.

\bibitem{Sakurai2020a}
J.~J. Sakurai and J.~Napolitano.
\newblock {\em Modern quantum mechanics}.
\newblock Cambrige University Press, 3 edition, 2020.

\bibitem{Quarteroni2011a}
A.~Quarteroni, G.~Rozza, and A.~Manzoni.
\newblock Certified reduced basis approximation for parametrized partial differential equations and applications.
\newblock {\em J. Math. Ind.}, 1:1--49, 2011.

\end{thebibliography}
\end{document}